# A Novel Pixel-Chip-Based Region-of-Interest Readout Circuit Design

Shi-Qiang Zhou[1], Li-Rong Xie[2], Dong Wang[1,*], Cheng Lian[1], Si-Ying Liu[1], Zi-Yi Zhang[1], Xiang-Ming Sun[1], Hong-Bang Liu[2], Chao-Song Gao[1], Jun Liu[1], Huan-Bo Feng[2], Di-Fan Yi[2,3]

[1] PLAC, Key Laboratory of Quark & Lepton Physics (MOE), Central China Normal University, Wuhan, 430079, China
[2] School of Physical Science and Technology, Guangxi University, Nanning 530004, China
[3] School of Physical Science, University of Chinese Academy of Sciences, Beijing, 100049, China

**Abstract:**

This paper presents a novel pixel chip readout scheme: the Region-of-Interest Readout Circuit (ROIRC), which is designed for large area, large array pixel chips and Gas Pixel Detector (GPD). This design employs a sentinel pixel detection strategy, enabling rapid identification and prioritized readout of the pixel regions containing signal events. During the scanning readout of these signal events, ROIRC employs a Block-based readout approach, effectively minimizing the readout of non-signal pixels. The functionality of ROIRC has been successfully implemented on both the ASIC and FPGA platforms. In the tests of the ROIRC, the pixel chip embedded in the GPD is capable of detecting low-energy X-rays in the range of 2-10 keV and supports multiple event readouts, and the pixel chip can read out photo-electron signal events with the count rate up to $15\text{k} \times \text{cm}^{-2} \times \text{s}^{-1}$.

**keywords**: ROIRC, Topmetal-L, LPD, pixel chip, X-ray

## 1. Introduction

POLAR-2 is the next-generation space station detection for the Chinese POLAR experiment, which is based on the same Compton scattering measurement principle as POLAR, but with an extended energy range and an order of magnitude increase in total effective area for polarized events. The Low Energy X-ray Polarization Detector (LPD) is one of the three payloads in the POLAR-2 experiment. LPD is specifically designed to observe the polarization of Gamma-Ray Burst (GRB) prompt emission in the energy range of 2-10 keV and to measure the polarization of the GRB as well as their very early X-ray afterglow, both in terms of polarization degree and polarization direction[2-4]. This observation is achieved using an array of X-ray photoelectric polarimeters based on GPD. Pixel chips have excellent characteristics in terms of energy resolution, fast time response, high spatial resolution. They have been widely used in space exploration; thus, the anode pixel readout chip is one of the core devices [5-10]. Pixel chips typically consist of two parts: the pixel array and the readout circuit.

The readout circuit significantly impacts the performance of the pixel chip. Currently, there have been a series of studies on the readout design of pixel chips. Typical readout designs of pixel chips, such as the ALPIDE chip, are applied in the ALICE ITS experiment at the Large Hadron Collider at CERN. The pixel chip readout of ALPIDE employs a hit-driven fashion, reading out only the pixels that are hit by particles. The in-pixel multiple-event is read out asynchronously by the priority encoder circuit in each double column. This design is not only fast in response but also power efficient, as the expected occupancy is low, and only hit pixels are read out in a hit-driven fashion[10-13]. In the Imaging X-ray Polarimeter Explorer (IXPE) and the Enhanced X-ray Timing Polarimetry (eXTP) mission, the XPOL-I and XPOL-III pixel chips are employed. These pixel chips possess self-triggering capability, enabling the localization of regions of interest that contain photo electron tracks. Within the chip, every 2 x 2 pixel array forms a trigger mini-cluster. When a mini-cluster is triggered by a signal event, the core logic automatically identifies all triggered mini-clusters and defines the the Region of Trigger (ROT) that contains them. By adding a pre-defined padding on the four borders—to define the region of interest (ROI) for the event capture and readout. In the XPOL-III, the approach to ROI definition has been upgraded, which delivers enhanced flexibility. This not only reduces the number of pixels within the ROI but also effectively shortens the readout time for signal events[11-18].

The electronics system for the cosmic X-ray polarization detection (CXPD), which functions as a prototype detector for the LPD, employs the Topmetal-II⁻ chip for its anode readout[19-20] and has achieved on-orbit capture of photo-electron signals. The successful operation of CXPD has determined that the Topmetal series of chips will be used as the anode readout for GPD in the LPD. The readout modules of these chips employ the rolling shutter scheme. This approach operates by sequentially reading out each pixel, irrespective of whether it has been hit by the signal event, and requires the waiting period of one frame time after each readout cycle before initiating the next operation [21-24]. Consequently, the frame refresh time increases with the size of the chip array. For the large-array Topmetal-M1 and M2 chips, the array is divided into 16 channels for parallel readout to enhance the readout rate; each readout channel requires the use of high-power analog buffers to drive signals, thereby leading to the significant increase in the overall power consumption of the chips.

Compared to CXPD, the pixel chips in LPD require key characteristics such as high effective area, high count rate, and low power consumption. Thus, large-array Topmetal pixel chips employing the Rolling Shutter readout mode struggle to meet the application requirements of the LPD. Consequently, a novel readout scheme needs to be developed based on previous generations of Topmetal chips.

This paper introduces a novel readout scheme: Region-of-Interest Readout Circuit (ROIRC), which consists of two components: the scanning module and the co-processing module. The scanning module serves as the readout circuit for the pixel chip Topmetal-L, which will be integrated into the LPD design. The co-processing module is implemented within the FPGA in the electronics. The design purpose of the ROIRC is to rapidly identify the pixels hit by signal events as priority readout areas, thereby reducing the waiting time for event readout. The ROIRC method for determining priority readout areas is implemented through the FPGA in the electronics, without the need for comparators to perform threshold comparisons, thereby simplifying the analog circuit architecture of the pixel unit. This approach meets with the LPD requirement for low-power design in pixel chips. This paper has tested the ROIRC readout scheme and successfully validated it.

## 2. Topmetal-L readout module design

Topmetal-L is a large-area CMOS pixel sensor chip designed for LPD. It is based on the Topmetal-II⁻ and Topmetal-M chips and fabricated using GSMC 130 nm CMOS process. As shown in the Fig.2, the total size of the Topmetal-L chip is 17mm x 24mm including a pixel matrix of 356 (row) x 512 (column) with the periphery circuit. The readout circuits are placed at the left and bottom of the pixel matrix. All the IO pads are located at the left, right and bottom of the chip to make it easy to be assembled in multi-chip applications. The total size of the each pixel sensor is 45 x 45 um² and the exposed noninsulated area is 26 x 26 um² . Each Topmetal is surrounded by a guard ring with the same metal layer, which is covered by an insulating layer. The coupling capacitance between the guard ring and the top metal can be used for pixel performance calibration. The guard ring of each pixel can be applied with an external voltage signal to emulate the electrons generated by particle hits.

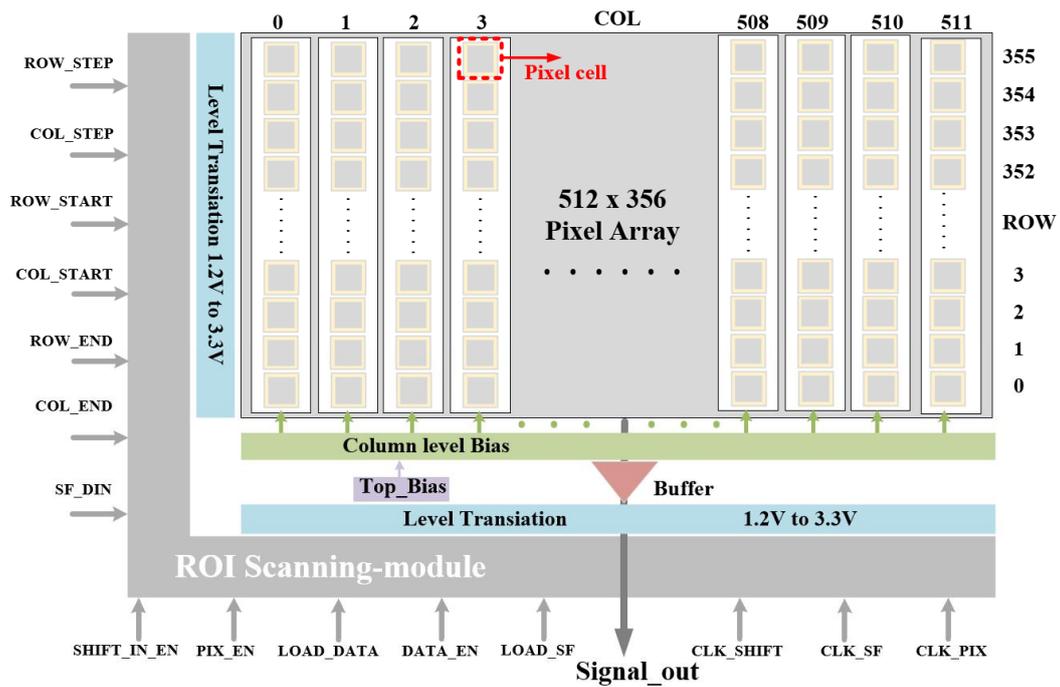

Fig.1. Analysis of the primary structural layout and distribution of Topmetal-L.

As shown in the Fig.3, each pixel unit consists of a Topmetal sensor, a charge sensitive amplifier (CSA), a two-stage source follower circuit, and row readout selection switches. The Topmetal working principle is based on a patch of the topmost metal layer acting as a charge collection electrode, placed in each pixel cell and the Topmetal is connected directly to the CSA. The Topmetal sensor of the pixel unit is responsible for collecting charge signal and converting it into voltage signal through the CSA for amplification.

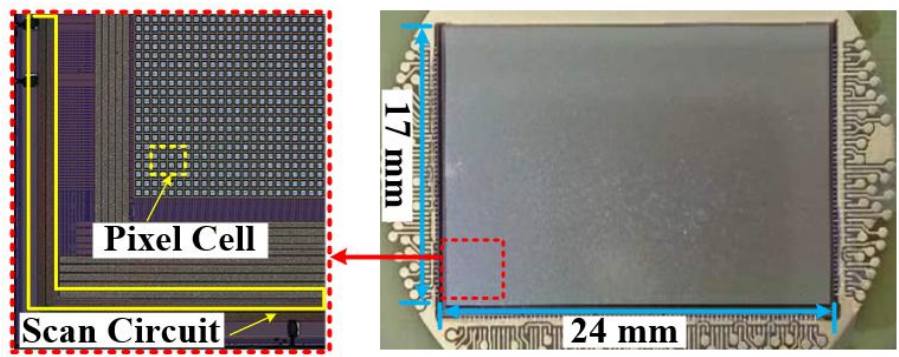

Fig.2. (a)Schematic representation of the local structure of the chip under the microscope. The lower left corner depicts the L-shaped readout circuit and level shifter circuit. The squares in the upper right represent individual pixel units. (b)Photograph of the entire pixel chip.

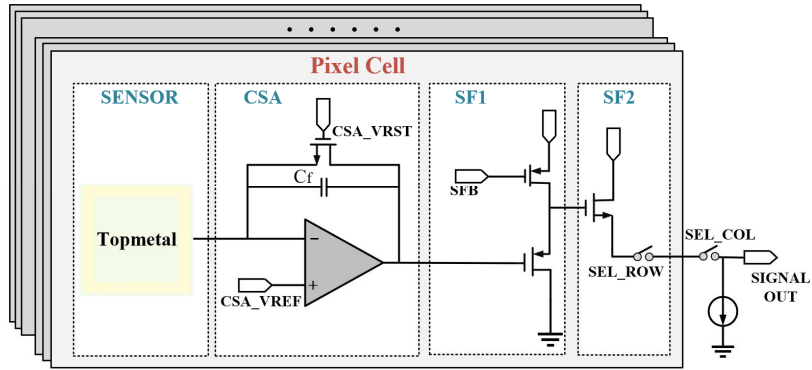

Fig. 3. Structure of a pixel unit cell in Topmetal-L.

The CSA consists of a folded cascode operational amplifier, a feedback capacitor $C_f$, and a discharge resistor $M_f$. The folded cascode architecture provides high gain and superior linearity, as shown in the Fig.4. The feedback capacitor $C_f$ (~1 fF), formed by the parasitic capacitance between two metal layers. The charge-to-voltage conversion gain of the CSA is $\triangle V_{out} / Q_{in} = -1 / C_f$, where $Q_{in}$ is the input charge, and $\triangle V_{out}$ is the output voltage. The decay time constant of the CSA output signal is given by $\tau = R_f \cdot C_f$, where $R_f$ represents the equivalent resistance of $M_f$. By adjusting the gate voltage of $M_f$, we can modify the value of $R_f$, thereby controlling the decay time of the output signal. Thus, the Topmetal chip employs the decay-based mechanism for pixel circuit reset. This design ensures that charge collection within each pixel remains continuous and independent of the operation of the row and column selection switches. However, this architecture requires that the row and column selection switches can rapidly and precisely locate the pixel region hit by the signal event.

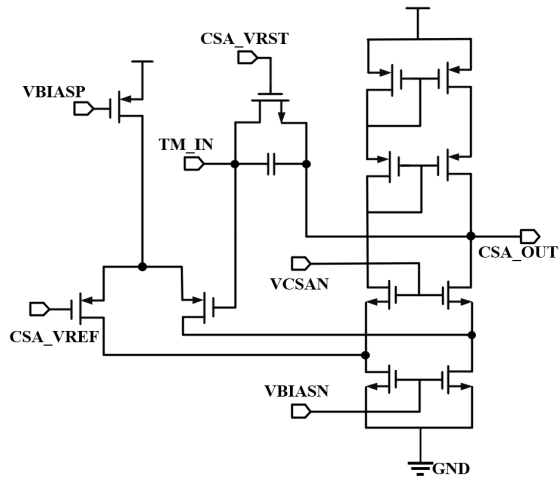

Fig. 4. Structure of the operational amplifier in the CSA.

As shown in the Fig.5, the single pixel response to a particle signal is depicted. Upon charge deposition, the sensor output exhibits a rapid step increase, followed by an exponential decay. The X-axis shows relative decay time, while the Y-axis indicates output signal amplitude. Time 1~5 represent readouts of the signal at different instances.

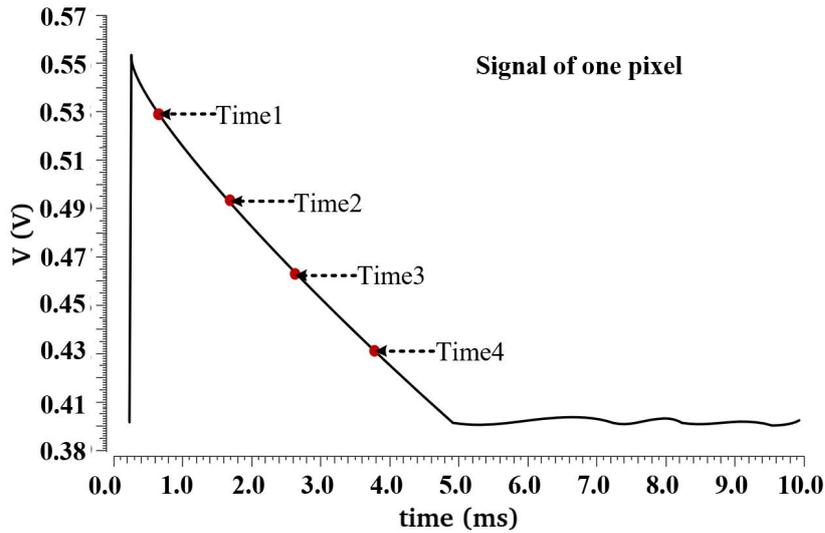

Fig. 5. The output response and decay characteristics of the CSA following charge collection at one single pixel, along with the cor responding energy amplitude obtained through signal sampling at different time points.

The CSA output is processed through two stages of the two-stage source follower. The signal is read out via row and column switches. In each column, all row readout selection switch outputs are connected to the column readout selection switch. Analog outputs from the pixel signal are transmitted out by analog buffer. The relevant parameters of the Topmetal-L chip are summarized in the Table 1. As detailed in [1], these parameters were both theoretically derived and systematically verified through measurement. The following discussion will focus on the readout circuit design of the chip.

Table 1. Key parameter of the pixel chip

| Parameters | Topmetal-II⁻ | Topmetal-M1 | Topmetal-L |
|---|---|---|---|
| Chip Size[mm$^2$] | 8 x 9 | 18 x 23 | 17 x 24 |
| Pixel Array | 72 x 72 | 400 x 512 | 356 x 512 |
| Pixel Density | 145 | 625 | 494 |
| Power Density | 2.778 | 0.826 | 0.195 |
| ENC[e-] | 13.9 | ~50 | 22.8 |
| Readout Mode | Rolling Shutter | Rolling Shutter | ROIRC |
| Readout channel | 1 x 1 | 1 x 16 | 1 x 1 |
| Pixel Gain | - | 78.6 | 76.04 |
| Frame rate | 2.6ms/frame | 1.2ms/frame | 0.7ms/frame |

The readout circuit of the Topmetal-L chip, serving as a critical component of the ROIRC i.e., the scanning module, is implemented using the standardized digital ASIC process. This L-shaped scanning module, embedded in the lower left corner of the pixel chip, manages data readout. Compared to the Rolling Shutter readout circuit, the advantage of this design lies in its ability to adjust its working method through multiple sets of parameter configurations, thereby offering flexibility and diversity in scanning schemes.

Table 2. Configuration Parameter Description

| Parameter | Description |
|---|---|
| CLK SHIFT | Scan parameter configuration clock, 50MHz. |
| CLK PIX | Pixel scan clock, 10MHz. |
| ROW START IN | Determine the row start position for pixel scanning. |
| ROW STEP IN | Determine the row step position for pixel scanning. |
| ROW END IN | Determine the row end position for pixel scanning. |
| COL START IN | Determine the col start position for pixel scanning. |
| COL STEP IN | Determine the col step position for pixel scanning. |
| COL END IN | Determine the col end position for pixel scanning. |
| DATA SHIFT EN | Scan parameter configuration enable. |
| LOAD DATA EN | Configuration parameter load enable. |

As shown in the Fig.6, the scanning module employs serial input of multi-bit data for each parameter configuration. The scanning logic, parameter configuration logic operate on independent clocks. This approach allows for customized design of different logic blocks to meet their specific timing requirements. The row readout selection switch of the pixel unit, the chip-level column readout selection switch, and the control of the column start switch are all connected by the scanning module. It achieves timing convergence under the 50MHz clock constraint for both the scanning logic and data configuration logic. However, the scanning logic clock frequency is also limited by the analog circuit, i.e., whether the analog signals of pixels before switching can be read out within the constrained scanning logic clock period. This is due to the limitation caused by the routing of large-array pixel chips, and we will focus on investigating this issue in subsequent chip iterations. In the ROIRC testing, the 10MHz scanning clock is utilized in pixel chip testing.

The function of each parameter is shown in the Table 2. The pixel switching time T_scan is 100ns, the data configure time T_shift is 20ns, the data configuration time is 20ns per bit. Each parameter configuration involves a 10 bit serial data and two enable load bits. The time required to complete one data configuration T_data is 220ns. Input signals for the scanning module are supplied by the co-processing module, whose circuit functions are implemented in an FPGA. This FPGA based design provides ROIRC greater flexibility and adaptability, enhancing compatibility with various low-energy X-ray detector electronic systems. The co-processing module design can also be ported to a digital ASIC implementation. The operational principles of the scanning module and the co-processing module, their working relationship, and the implementation scheme are illustrated in the Fig.6.

As summarized in Table 1, compared to the similar pixel array in the Topmetal-M chip, the Topmetal-L requires only a single readout channel to implement the ROIRC readout logic, reducing the number of analog buffers to one-fifteenth of the original. Simulation results show that the operating current of the analog buffer in a single readout channel is 10 mA, which effectively reduces the power consumption of the chip. This design enables each chip to utilize only one analog-to-digital converter, facilitating the low power integration of the detector system. The low-power design of the chip at the analog circuit level is described in [1].

This section primarily focuses on the circuit design from the perspective of the ASIC. In the following section, we will present a comprehensive overview of the ROIRC design, integrating the FPGA within the electronics system.

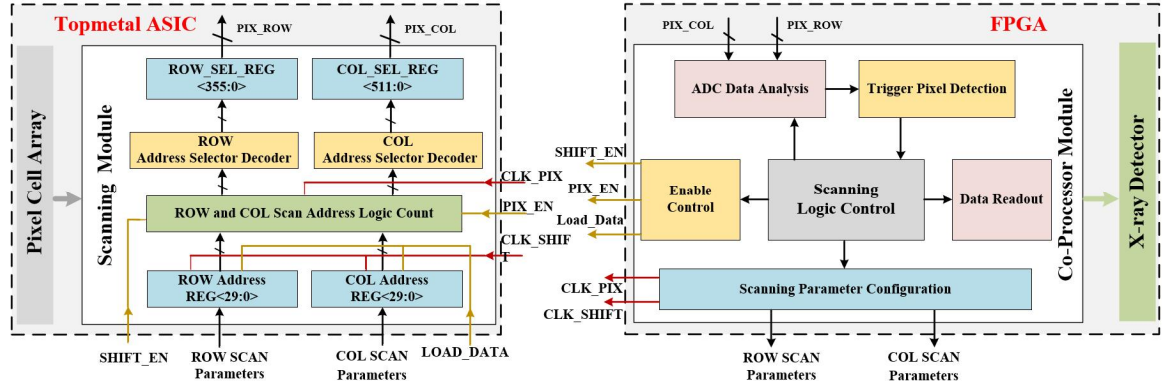

Fig. 6. Design principle of the ROIRC structure. It is implemented by the on-chip readout ASIC and FPGA in the electronics system.

## 3. Region of interest readout

The ROIRC comprises two core components: the scanning module integrated into the Topmetal-L chip and the co-processing module implemented in the FPGA. The scanning module receives parameters from the co-processing module, updates the scanning configurations, and reads out the pixels that have been scanned. The co-processing module is responsible for determining the scanning method and for sending control parameters to the scanning module. The scanning area is determined by setting the pixel start address, pixel end address, and the number of row and column steps in the pixel array.

The ROIRC workflow comprises three sequential phases: 1. In the initial phase, the co-processing module instructs the scanning module to perform readout at uniform row and column intervals. This process is defined as ``sentinel monitoring scanning'', where the pixels scanned in this process are defined as ``sentinel pixels''. Those sentinel pixels that are hit by signal events are defined as ``trigger pixels''. 2. The scanning module calculates the target area to be scanned based on the positions of these ``trigger pixels''. 3. To ensure complete readout of the signal events, the ROIRC performs peripheral expansion of the area that needs to be read out; this process is defined as ``inflation processing''. This paper elaborates on three core components of the ROIRC algorithm.

As illustrated in the Fig.7, the schematic diagram illustrates the random selection of rows 123--136 and columns 51--71 from the chip matrix. The scanning module divides the parameters received from the co-processing module into six distinct groups. With these six sets of parameters, the working method of the scanning module can be determined. Fig.7 illustrates the process by which the ROIRC captures and records the complete particle trajectory.

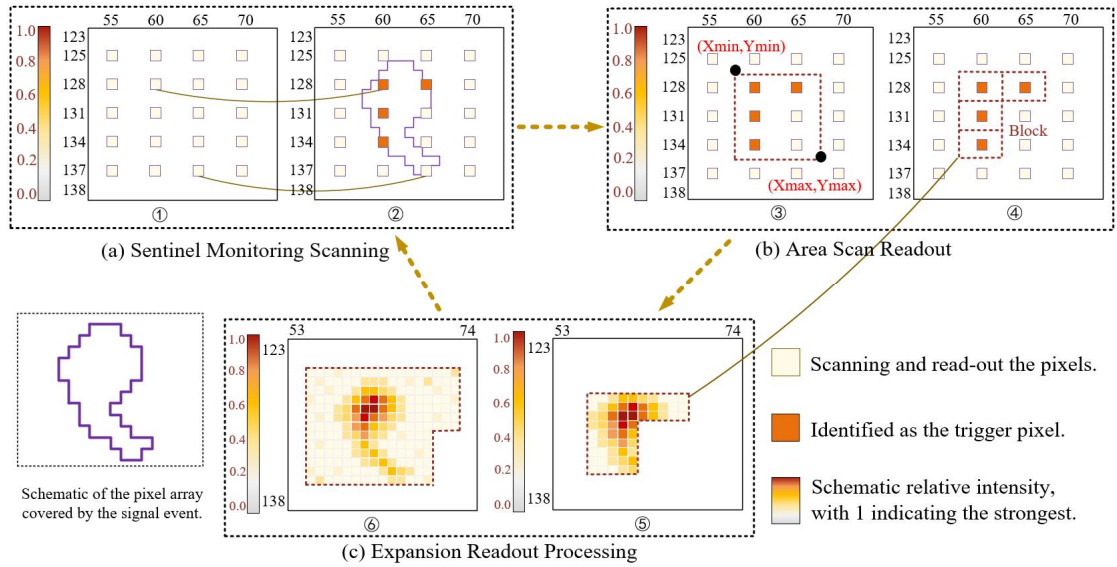

Fig. 7. ①The process of sentinel monitoring scanning. ②ROIRC determines the arrival of signal events through threshold comparison. ③The readout method employing the calculation of the minimum rectangle. ④The readout method employing the block approach. ⑤The scanning module performs readout based on the block containing the trigger pixel. ⑥Region scanning following dilation processing

## 3.1 Sentinel monitoring scanning and threshold comparison

The sentinel monitoring scanning process is illustrated in the Fig.8. Its purpose is to quickly determine the arrival of particle events by increasing the frame rate. Sentinel pixels are uniformly distributed across rows and columns of the pixel array, the distribution interval of the sentinel pixels is determined by the row/column step parameters, and only these pixels are read out during sentinel scanning. The co-processing module stores the data from each sentinel scan frame. If no signal event is detected, the scanning module repetitively executes this process.

The co-processing module stores the data from each sentinel scan frame. During subsequent frame scans, the current sentinel pixel data are subtracted from the corresponding values of the previous frame. If the difference exceeds the set threshold, it is determined that an effective signal event has occurred around that sentinel pixel, and the pixel is identified as the trigger pixel. The co-processing module records the address information of all trigger pixels in the current frame.

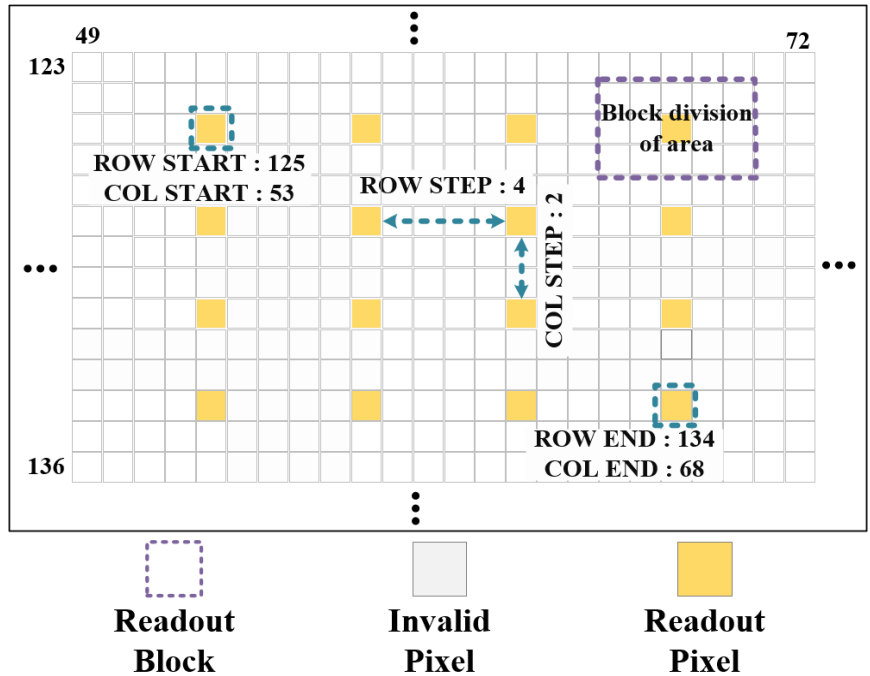

Fig.8. Parameters configuration for sentinel scanning mode.

The time T_sen required for sentinel monitoring scanning per frame can be calculated using the equation 1, where N_pixel is the total number of pixels in the array, N_r-step is the number of row steps, and N_c-step is the number of column steps during the scanning process. T_scan, T_data are known (see more details in the section 2. )

**3.2 Region scanning**

The region scanning scheme is executed upon the detection of the trigger sentinel pixel, as shown in Fig.9 This scheme employs a block-based scanning method, where each triggered sentinel pixel corresponds to a predefined block region, with the block's position determined by the location of the triggering sentinel.

For all trigger sentinels within a cluster event, the ROIRC divides block regions based on the mutual positions of trigger sentinel pixels: adjacent sentinels are assigned blocks with boundaries strictly defined by the preset row and column interval, while sentinels identified on the event periphery can have their block dimensions dynamically adjusted by the co-processing module. This strategy integrates adjacent blocks into a continuous readout region, thereby avoiding redundant pixel scanning while ensuring complete event coverage.

During the region scanning process, the ROIRC reads out pixel information from each block sequentially in row-column order. As shown in Fig.9 and Fig.19b, compared to the method of calculating the minimum bounding rectangle (i.e., determining the coordinates of X_max, X_min, Y_max, Y_min for the area), the block-based approach reduces both the number of pixels to be read out and the associated readout time, particularly for irregular or elongated tracks.

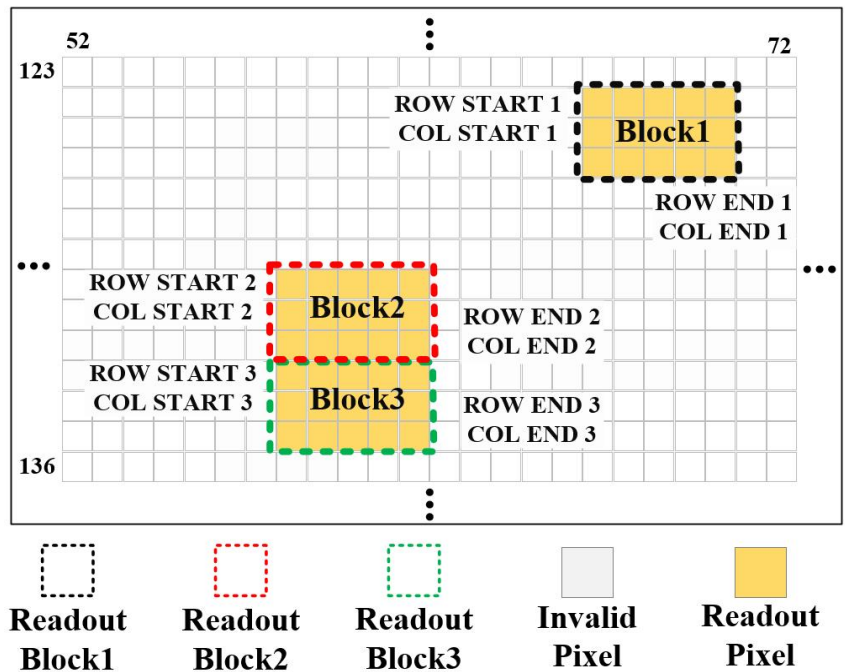

Fig.9.Parameters configuration for region scanning mode.

As shown in Fig.9, due to the large size of the chip's pixel array, it is possible that multiple signal events hit and are detected within the duration of a single sentinel scan frame. When multiple signal events are detected within the same sentinel scanning frame and are determined to be spatially distant from one another, the system assigns each event its own dedicated block scanning region. They are not combined into a single rectangular area.

The spacing between sentinel pixels can be flexibly configured according to the application requirements to achieve an optimal balance between frame rate and signal event capture probability. Once the region scanning is completed, the ROIRC returns to the sentinel monitoring scan mode to await subsequent signal events. Since the scanning module requires parameter reconfiguration before each block scan, ; thus, the single-block scanning time T_block is calculated using equation 2, where N_block is the number of pixels in a single block.

### 3.3 Dilation processing

As shown in the Fig.10, particle signal events typically deposit energy in irregular, spatially stochastic patterns across the pixel array. When edge signals exhibit insufficient amplitude, the difference between the sentinel frames before and after the event may not change significantly. This prevents the sentinel from being triggered as a trigger pixel. Consequently, reading out only the blocks containing the trigger pixels may result in the incomplete readout of the signal event. To address this issue, an inflation process is introduced during the region scanning to ensure complete signal-event readout. The concept of inflation processing is as follows: not only are the blocks containing the trigger pixels read out, but also the blocks surrounding the trigger pixels that contain sentinels are read out to ensure the complete readout of the signal event. The Fig.10 illustrates the implementation process of the inflation algorithm.

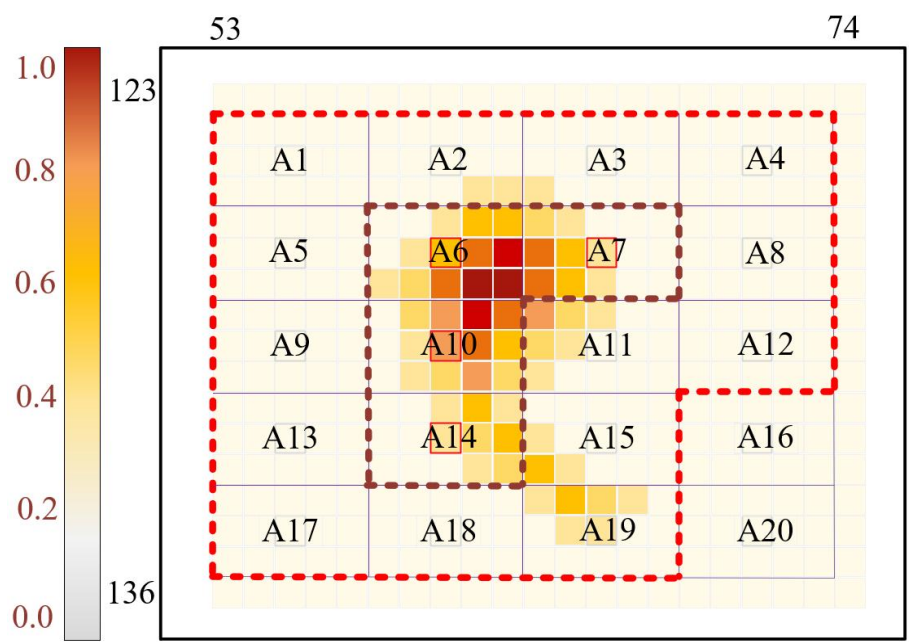

Fig.10. Schematic of region scanning expansion. The black lines indicate the block region where the block sentinels are located, while the red lines show the expanded region.

## 4. Test results

This section presents the test results of the Topmetal-L chip integrated into the GPD and operated under the ROIRC readout logic. The readout architecture of the pixel chip in the LPD detector must read-out complete photo-electron track imaging, high count-rate capability, and rapid readout of the pixel region hit by signal events. Consequently, the evaluation focuses on the readout mechanism, data throughput, integrity of signal events, and the acquisition of particle tracks with varying trajectories.

It should be emphasized that the Topmetal-L chip and the ROIRC readout scheme constitute only one part of the overall detection system. The detector performance is influenced by multiple factors, including the internal components of the GPD, the properties of the gas mixture, and the electronic system. Therefore, all tests described hereafter were conducted under the operating conditions for both the chip and the detector, as defined in [1]. This ensures that the ROIRC logic not only achieves quickly readout but also maintains the reliability and validity of the data acquired and read out by the detector. In the section \ref{sec.4.1}. details the setup of the testing platform.

**4.1 Test setup**

As shown in the Fig.11. The GPD is divided into three functional regions: the electron drift region, the electron multiplication region, and the charge collection region. The gas microchannel plate (GMCP) serves as the electron multiplier in the electron multiplication region. The GMCP has the diameter of 25 mm, the thickness of 400 um, the pore diameter of 50 um, and the pore pitch of 60 um arranged in the triangular pattern.

The drift gap between the cathode and the GMCP is 14 mm, and the induction gap between the GMCP and the anode is 4 mm. The detector was operated with the following voltages: the cathode voltage (V-drift) at -3600 V, the GMCP top surface (V-top) at -1840 V, and the GMCP bottom surface (V-bottom) at -600 V, while the anode was grounded. The corresponding electric field strengths in the three regions are 1.76 kV/cm in the electron drift region, 41.3 kV/cm in the electron multiplication region, and 1.5 kV/cm in the charge collection region. The working gas within the chamber consists of helium and dimethyl ether (DME) in a 3:7 ratio, as defined in [1]

X-rays will be emitted by the $^{55}Fe$ source or the X-ray generator. These X-rays pass through the beryllium window on the gas chamber to enter the chamber. The mixed gas inside the chamber enhances the interaction between the X-rays and the gas molecules in the electron drift region, effectively inducing photoelectric effect and converting X-rays into photo electrons. Under the influence of the electric field, these electrons move to the upper surface of the GMCP and enter the channels through the small orifices. Within these channels, the cascade multiplication of the initial electrons occurs, leading to the generation of a significant number of secondary electrons. These secondary electrons eventually exit the GMCP and are collected by the Topmetal. X-rays will be emitted by the $^{55}Fe$ source or the X-ray generator. These X-rays pass through a beryllium window on the gas chamber to enter the chamber.

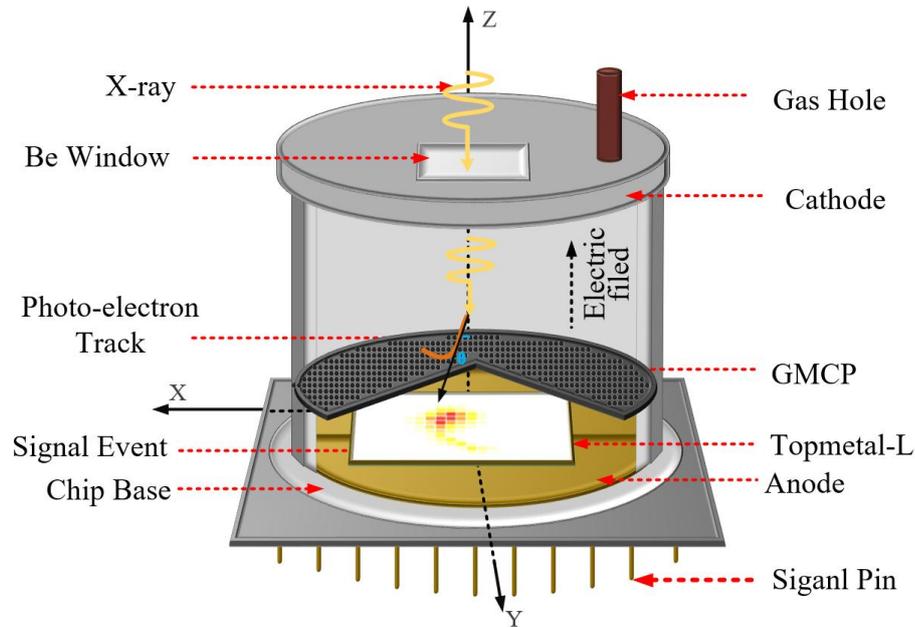

Fig.11. The internal structure and working principle of the GPD.

The test electronics comprise the bonding board, the readout board, and the FPGA core control board, with their interconnections shown in Fig.12. The readout board serves two primary functions. Firstly, it utilizes an onboard 8-channel DAC8568 chip to supply adjustable bias voltages to the Topmetal-L chip. By configuring the different DAC channels, key performance parameters such as the decay time constant can be precisely tuned to ensure stable chip operation. Secondly, the board transmits the pixel readout signals to the ADC analog-to-digital converter located on the FPGA core control board. This ADC features an input dynamic range of –1V to +1V, the sampling rate of 40 MSPS, and the sampling resolution was 12 bits; thus, the minimum scale of the ADC was approximately 0.49 mV.

The FPGA core control board implements the co-processing module and handles data transmission and processing tasks. The exchange of control commands and data transmission between the system and the PC are both accomplished via the PCIe Express bus interface. Furthermore, the readout board can be connected to the external oscilloscope for real-time monitoring of pixel output signals. Test operators may also inject square-wave signals into the guard-ring structure of the Topmetal-L chip to simulate the physical process of negative charge injection into the top-metal layer.

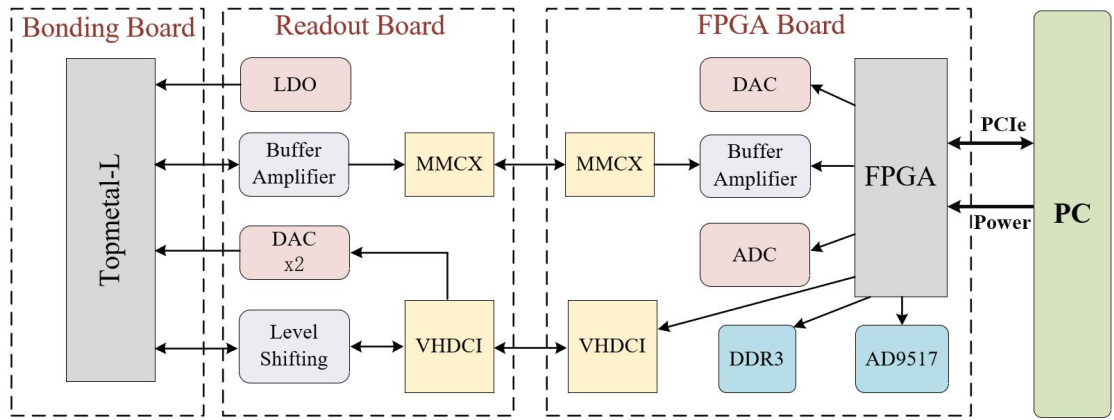

Fig.12. Structure of the electronic system.

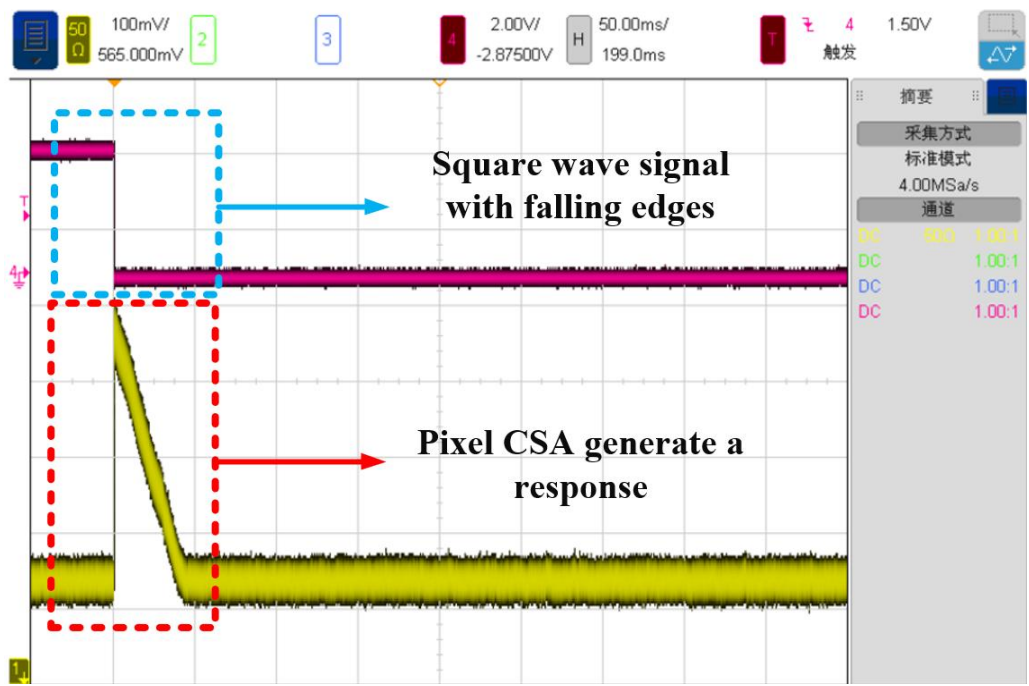

Fig.13. The yellow signal represents the CSA output of the pixel chip, and the pink signal line indicates the injected square wave sig-nal. Each pixel responds to the falling edge of the square wave signal and decays slowly.

**4.2 Noise testing and pixel shielding**

As described in section \ref{sec.2.1}, the condition of "trigger pixels" is related to the trigger threshold set in the co-processing module. Due to the limitations of

CMOS technology, some pixels in the chip may exhibit higher noise levels, as shown in Fig.14. If these high-noise pixels are selected as sentinel pixels, these noise fluctuations can exceed the set threshold, causing the co-processing module to trigger erroneously and leading to frequent initiation of region scanning.

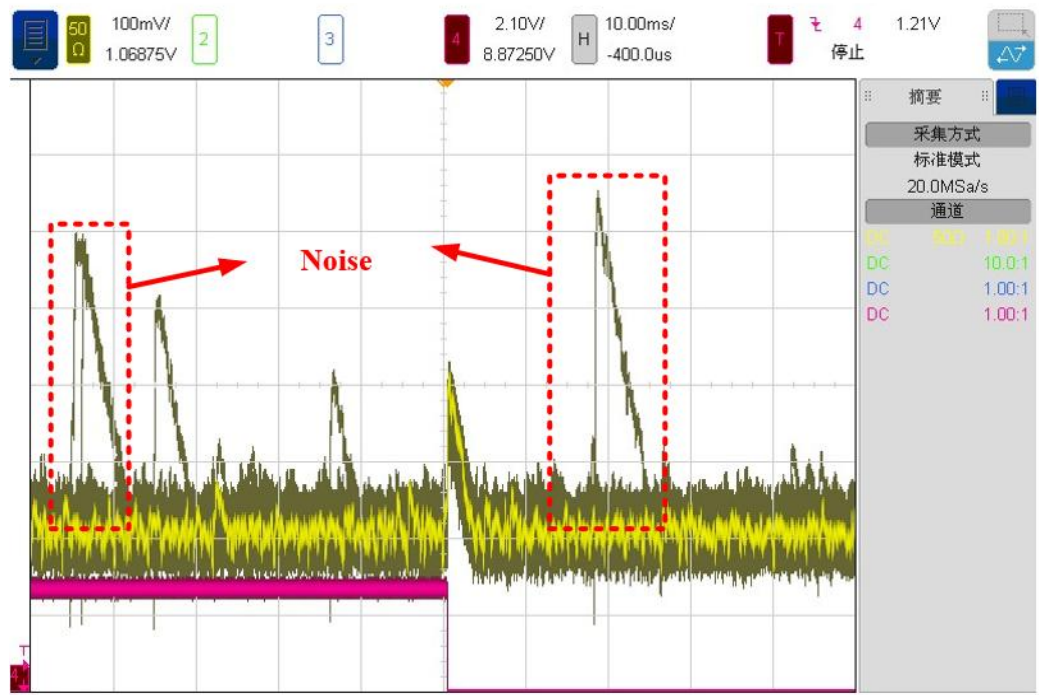

Fig.14. Display of effective charge signal and noise signal in the single pixel.

In this test, we compared the output signal baselines of all pixels in the Topmetal-L chip across consecutive frames and visualized in binary the pixels with differences exceeding the threshold, as shown in Fig.15a. As the threshold increases, the number of white points (pixels with differences exceeding the threshold) decreases. Setting the threshold too high can mask "defective pixels", but this will lead to the loss of valid signal events, especially during low-energy X-ray detection. To address this issue, this paper proposes implementing the bad-pixel masking algorithm. ROIRC records and masks the addresses of these defective pixels. Even if such pixels serve as sentinel pixels, they will no longer participate in the sentinel triggering logic. In this experiment, with no signal injected into the chip, pixels with the difference between the current and previous frames exceeding the 100 mV threshold were masked. Fig.15b shows the binarized image after masking, demonstrating that the bad-pixel masking strategy effectively suppresses false triggering.

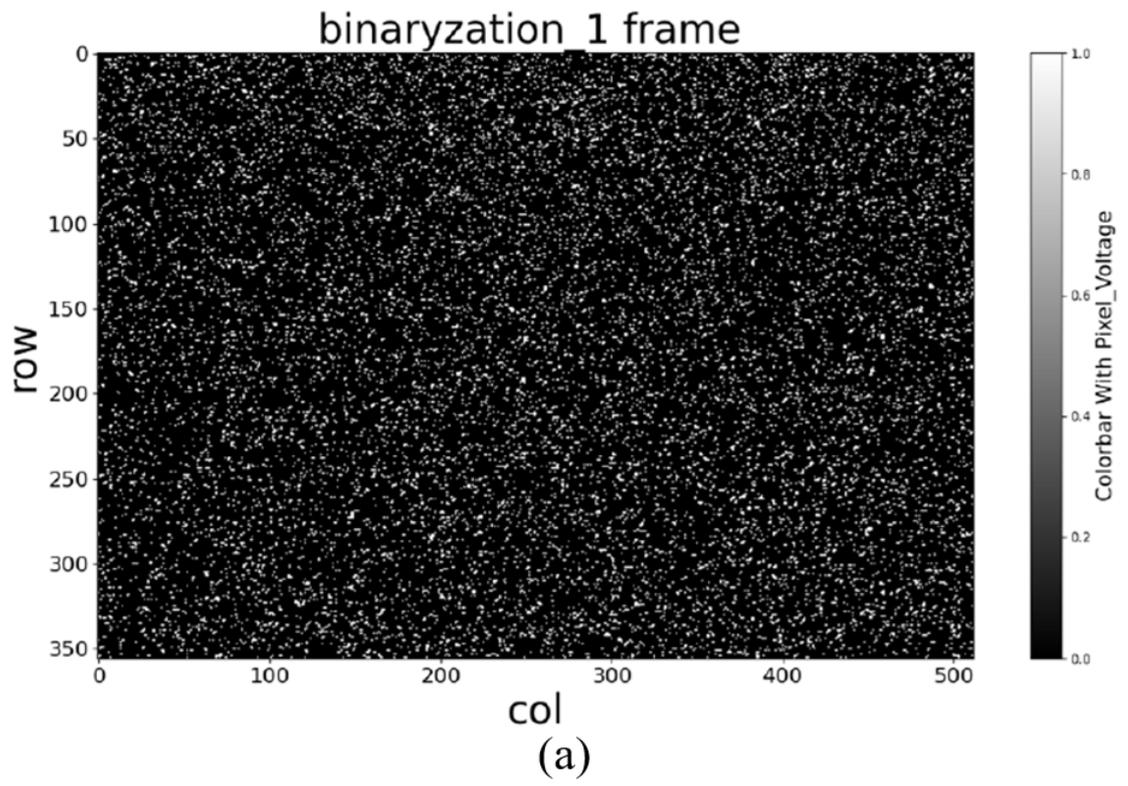
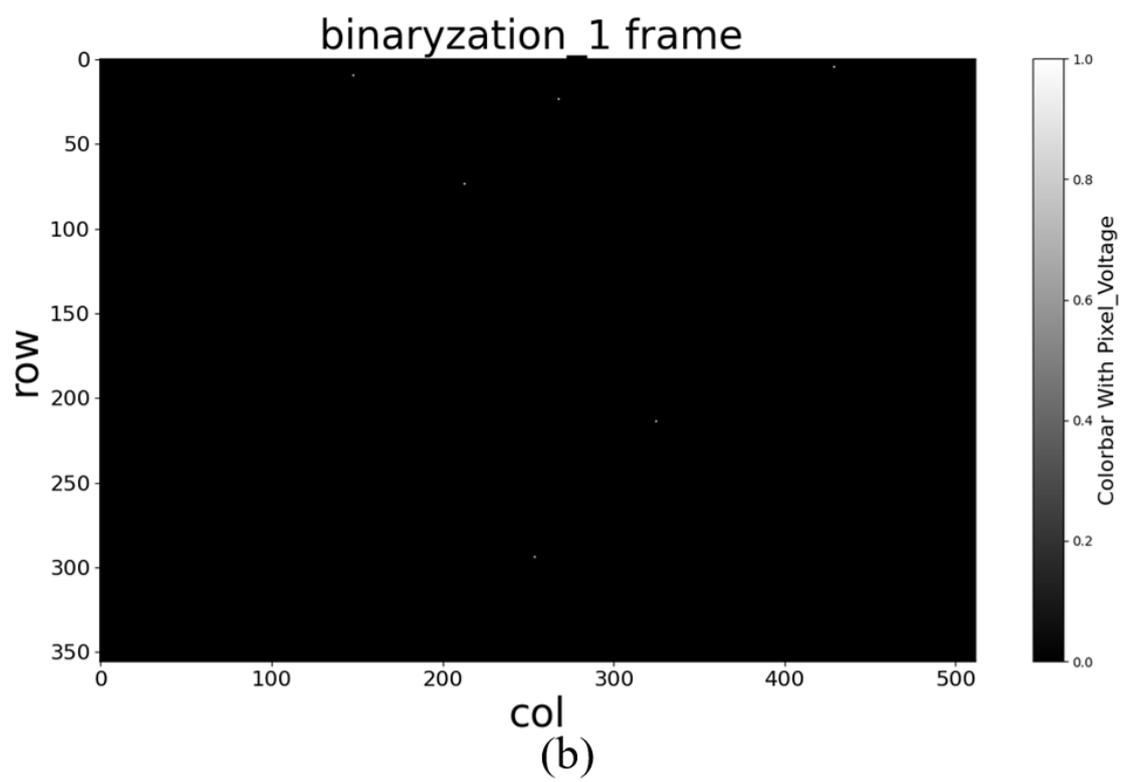

Fig.15. Results of high-noise pixels.(a)Without shielding.(b)With shielding.

## 4.3 Validation of readout scheme feasibility

The ROIRC scheme detects signal events through sentinel pixels. The spacing interval between these sentinels is a critical parameter, directly impacting the system's trigger efficiency and frame rate. Operators can configure ROIRC parameters based on the energy of signal events and their spatial distribution across the pixel array. We therefore performed Monte Carlo simulations, which revealed that with a row and column spacing of 5 pixels and a trigger threshold of 200 ADC counts (as established in Section 4.2 for bad pixel filtering), the chip achieves a collection efficiency exceeding ~90% for 3 keV photo-electron signals. Consequently, the following parameter set was adopted for ROIRC in this experiment: the scanning module's start address for rows and columns is set to 0, with end addresses at row 355 and column 511; the step size for both rows and columns is 5 pixels; each block corresponding to a sentinel pixel contains 25 pixels; the minimum region scanning size (i.e., triggered by a single sentinel pixel, referring to the extent of the expanded region) covers 361 pixels; and the trigger threshold is set to 200 ADC counts.

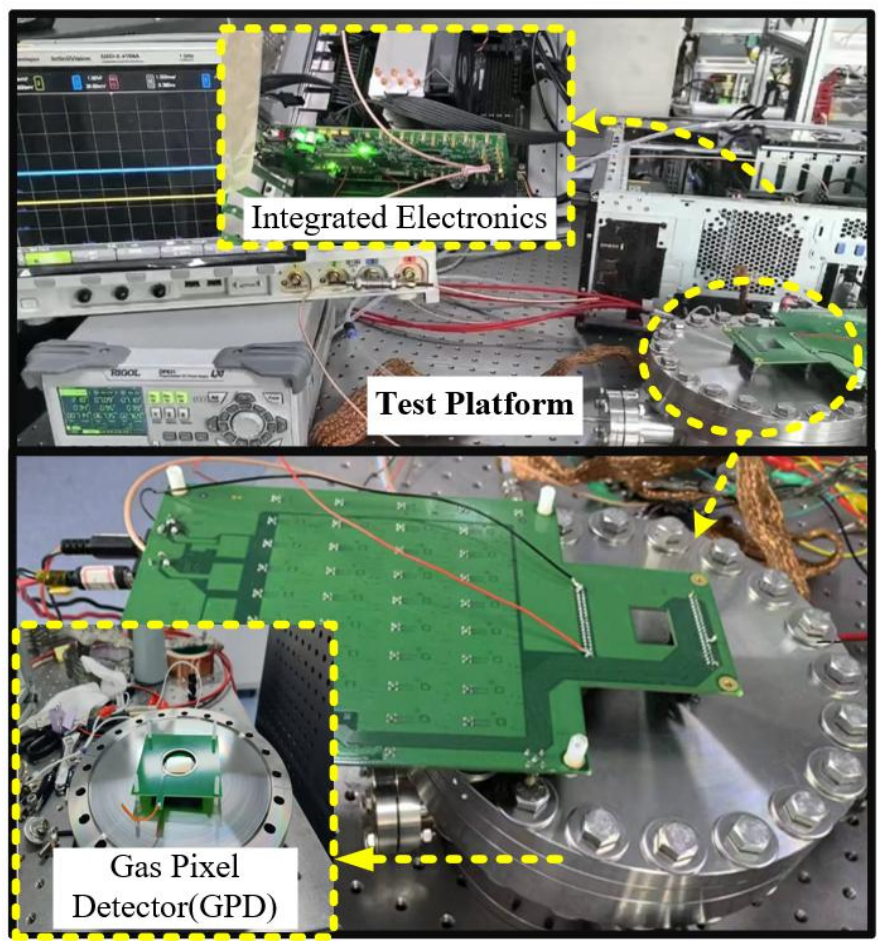

Fig.16. 55Fe source test platform.

We utilized the radioactive source $^{55}Fe$ for the experiments in this section. The radioactive source $^{55}Fe$ emits monoenergetic 5.9 keV X-rays and is used to assess the detector's capability to record and reconstruct the photoelectron tracks generated by low-energy X-rays under the ROIRC readout scheme.

The Fig.17 illustrates the capture of signal events by ROIRC. The Fig.17a① displays the sentinel scanning process in the 512 x 356 pixel matrix, where the evenly spaced blue dots represent the pixel values read out by ROIRC during sentinel monitoring scanning, with all other pixel values set to 0, and the red wireframe enclosing the triggered scout pixels. Fig.17a② represents the event signals from the regional readout, while Fig.17a③ provides a magnified view of these event signals.

Fig.17b shows the sentinel pixels information at the local position of the pixel chip. Each readout sentinel pixel is evenly spaced and uniformly distributed. The scanning region in Fig.17b② corresponds to the position of the triggered scout pixel in Fig.17b① .

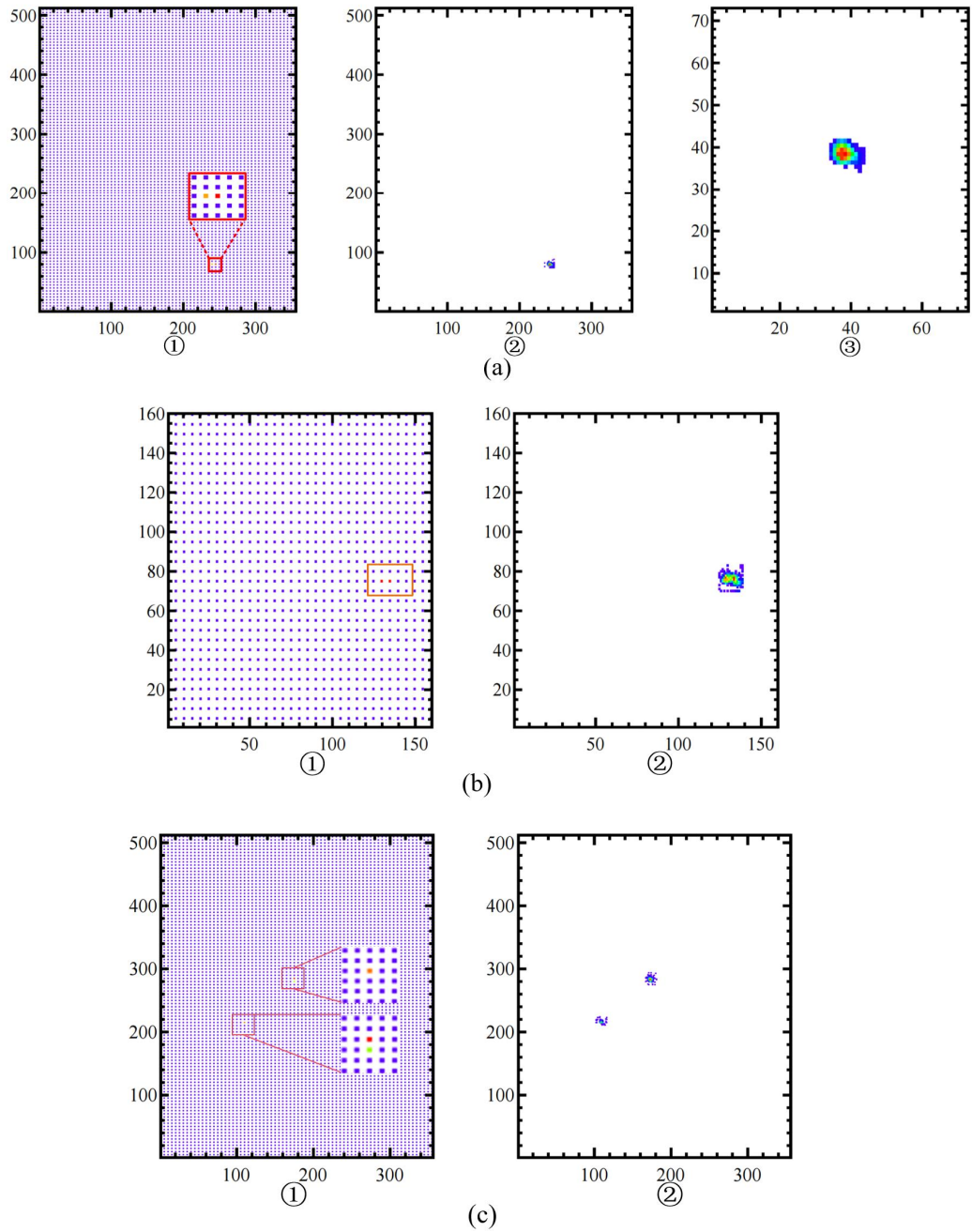

Fig.17. Validation of ROIRC logic. Blue pixels represent pixel baseline noise, and pixels with colors closer to red indicate higher energy deposition. The red wire frame marks the detailed display area.

The Fig.17 demonstrates that in one single frame of sentinel monitoring scanning, ROIRC can support the triggering of multiple sentinels and perform multi-region readout operations based on the sequence of trigger pixel positions. According to equation.1, equation.2 and equation.3, under the scanning step of 5 pixels in both row and column directions, the maximum dead time required for reading effective event signals is 709.24 us. In contrast, under the Rolling Shutter readout mode, the maximum dead time reaches 18227.2 us. Because of this scanning scheme, the frame rate depends only on the chip's pixel cells and its highest scanning frequency.

**4.4 Complete Readout Rates**

ROIRC is capable of rapidly reading out event signals while ensuring that the signal events are entirely located within the pixel region of the region scanning. In the Fig.18, pixels marked in red indicate the triggered sentinel pixels. Address information of the sentinel data enables the back-end data processing to determine the scanning region. In the readout logic, all sentinel pixels are positioned at the center of the block region.

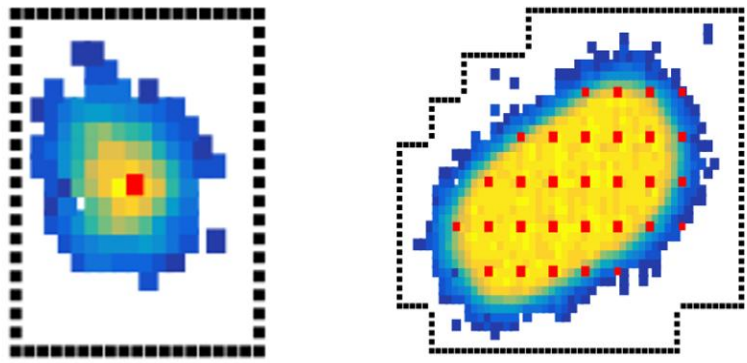

Fig.18. With the sentinel address as the center, the scanning region is identified. Red dots denote the positions of sentinel pixels, and the black dashed line indicates the region's boundary

By locating sentinel pixels, the data processing can determine whether the entire trajectory of the signal event is within the scanning region, thereby confirming whether the signal event has been fully read out. According to the internal configuration of the GPD, the gas mixture possesses a natural level of radioactivity owing to the presence of the small amount of radon (Rn), which decays and emits high-energy alpha particles. We conducted the comprehensive analysis of the integrity for both X-ray and alpha signals. The extended tracks produced by alpha signals are more spatially expansive, thereby serving as a more effective means for verifying the completeness of track readout within the region scanning procedure. The experiment collected approximately 10,000 signal events, and the validation results showed that the trajectories of all events were fully read out.

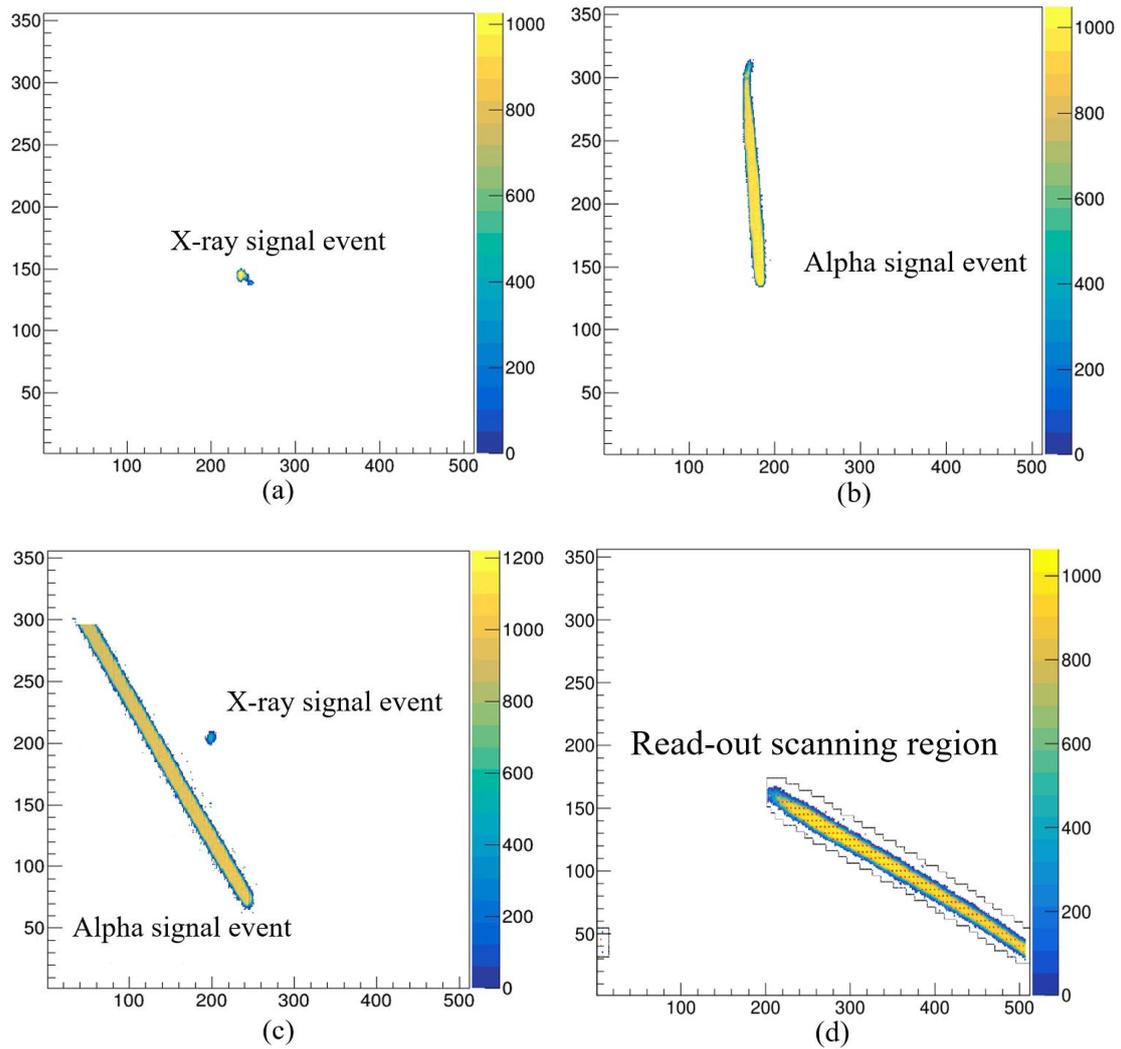

Fig.19. The ROIRC reads out the tracks of particles with different shapes in the Block-based measure.

ROIRC uses the block-based readout to capture signal events of various track shapes and pixel coverage. Even if multiple event signals are triggered in one single sentinel scanning, the readout logic ensures all event readout during the subsequent regional scanning. Fig.19 shows that Fig.19a presents particle tracks from $^{55}Fe$ signals, and Fig.19b shows tracks from Alpha particle signals. Fig.19c illustrates the simultaneous generation and reading of both signals during the same sentinel scanning period. As indicated in Fig.19d, the black wire frame indicates the pixel area (composed of all sentinel pixels' blocks) during region scanning.

### 4.6 Readout rate

One of the primary goals of the ROIRC design is to make the low-energy polarization detector (LPD) equipped with a large-array pixel chip capable of meeting the requirements of astrophysical observations. For instance, the peak photon count rate of GRB 221009A, the brightest GRB observed to date, reached about 4500 photon $\times cm^{-2} s^{-1}$ in the 2–10 keV band within a one-second interval [28]. Considering the detection efficiency of the GMCP is approximately 0.1, the signal event rate to be processed by the corresponding front-end electronics system is no less than 450 counts $\times cm^{-2} s^{-1}$ [1][3].

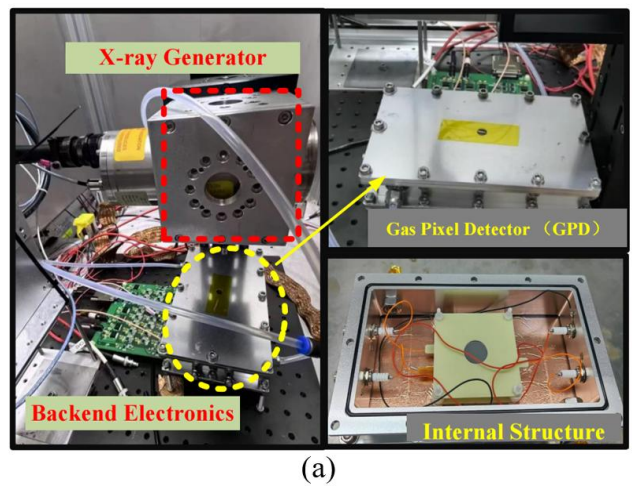

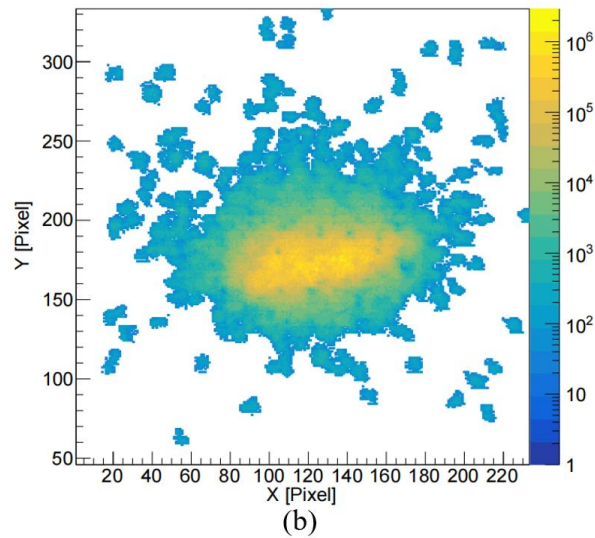

Fig. 20. (a) X-ray generator test platform; (b) Statistical analysis of photon energy and range per square centimeter per second

As described in section \ref{sec.2}, the signal amplitude acquired by the Topmetal-L pixel chip decays over time, and failure to complete the readout within this decay time window will result in signal loss. Therefore, the signal event count rate of the detector in the ROIRC readout mode needs to be evaluated, which refers to the number of events that the chip can fully process per unit time by the chip in this readout scheme.

In this counting experiment, the bias voltage settings for the Topmetal-L chip were consistent with those in [1]. Based on combined simulation and experimental results, the typical discharge rate of the CSA for charge signals under this system configuration is approximately 37.28 mV/ms.

The test results are shown in Fig.21 The Y-axis is defined as the number of signal events collected and read out by the Topmetal-L chip embedded in the GPD. The X-axis is defined as the relative intensity setting of the X-ray generator.

Under the ROIRC readout mode, when the signal events count rate is less than $15k \times cm^{-2} \times s^{-1}$, the chip count rate scales linearly with the increasing photon output from the X-ray generator, meaning there is no signal event overlap. However, when the signal event count rate exceeds $15k \times cm^{-2} \times s^{-1}$, the count rate no longer increases linearly, and some events will overlap. These results confirm that the ROIRC operated detector achieves a maximum usable readout rate of $\sim 15k \times cm^{-2} \times s^{-1}$, exceeding the 450 counts $\times cm^{-2} s^{-1}$ requirement set by the brightest known GRB scenarios.

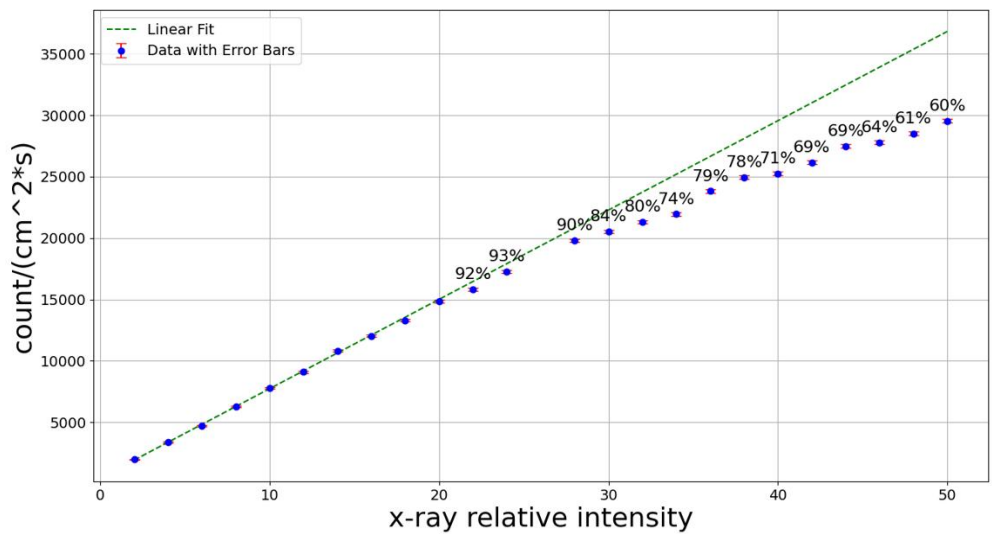

Fig. 21.   The maximum flux test of effective photons.

# 5. Conclusion

We have designed and tested a novel readout circuit design based on the pixel chip Topmetal-L, named ROIRC, which is implemented by the digital ASIC and the FPGA in the electronics. We have detailed its design structure, logical working principle, and functional behavior. Additionally, we have introduced test results, which include X-ray tests based on the radioactive element $^{55}Fe$ and the X-ray generator. ROIRC can be divided into two parts: the scanning module and the co-processing module. The scanning module, serving as the readout circuit for Topmetal-L, is primarily responsible for receiving parameters sent by the co-processing module, updating the scanning method, and reading out the scanned pixels. The main function of the co-processing module is to determine the scanning method and send control parameters to the scanning module. The entire algorithm achieves rapid and complete readout of signal events through sentinel monitoring scanning and regional scanning.

Under the ROIRC operating mode, GPD can detect low energy X-ray in the range of 2 to 10 keV. Multiple regions can be triggered and read out in one single frame of sentinel scanning detection. The block region readout method can effectively reduce the scanning of invalid pixels. When the effective photon count rate is less than $15k \times cm^{-2} \times s^{-1}$, the readout rate of ROIRC for signal events increases linearly. Due to the flexibility of ROIRC, its operating modes are diverse, making it particularly suitable for scenarios where pixel chips require large scale assembly, low power consumption, and large detection areas.